# Energy Stability in High Intensity Pulsed SC Proton Linac

A. Mosnier, CEA-DAPNIA, Gif-sur-Yvette, France

*Abstract*

Spallation sources dedicated to neutron scattering experiments, as well as multi-purpose facilities serving several applications call for pulsed mode operation of a high intensity proton linac. There is general agreement on the superconducting technology for the high-energy part, which offers some advantages, like higher gradient capabilities or operational costs reduction, as compared to room-temperatures accelerating structures. This mode of operation however could spoil the energy stability of the proton beam and needs thus to be carefully studied. First, transient beam-loading effects, arising from the large beam phase slippage along a multi-cell cavity and associated with the finite RF energy propagation, can induce significant energy modulation with a too small cell-to-cell coupling or a too large number of cells. Second, due to beam phase slippage effects along the linac, energy spread exhibits a larger sensitivity to cavity fields fluctuations than relativistic particles. A computer code, initially developed for electron beams has been extended to proton beams. It solves the 6xN coupled differential equations, needed to describe cavity fields and beam-cavity interactions of an ensemble of N cavities driven by one single power source. Simulation examples on a typical pulsed proton linac are given with various error sources, like Lorentz forces or microphonics detuning, input energy offsets, intensity jitters, etc...

## 1 INTRODUCTION

With the aim of studying the energy stability in a high intensity pulsed superconducting proton linac, which could be spoiled by transient beam loading and beam phase slippage effects, we used the two computer codes MULTICELL and PSTAB. The former[1], based on a multi-mode analysis, calculates the systematic energy modulation generated within a multi-cell cavity due to the finite speed of the rf wave propagation; the latter, initially developed for relativistic beams[2] has been extended to low beta beams and can handle all major field error sources (Lorentz forces, microphonics, input energy offsets, beam charge jitter, multiple cavities driven by a single power source, etc) including feedback system and extra power calculation. Due to lack of space, this paper is a shortened version of a longer report[7] and presents a few results of simulation for the case of a typical neutron spallation source, like ESS[3]. The relevant parameters of the SC High Energy Linac, which have been chosen for this study are shown in Table 1.

The bunch frequency during beam-on time is assumed, after funnelling of two beams emerging of 352.2 MHz RFQs, to be equal to the 704.4 MHz rf frequency of the SC cavities. Three different cavity types have been selected from input to exit of the linac and their energy gains as a function of the incoming energy are plotted in Fig.1 (synchronous phase of -30° included).

Table 1: Typical SC High Energy Linac for ESS

| | |
|---|---|
| Input energy | 85 MeV |
| Exit energy | 1.333 GeV |
| Peak beam current | 107 mA |
| Chopper duty factor | 60 % |
| Bunch train period | 600 ns |
| Number of bunch trains | 2000 |
| RF frequency | 704.4 MHz |

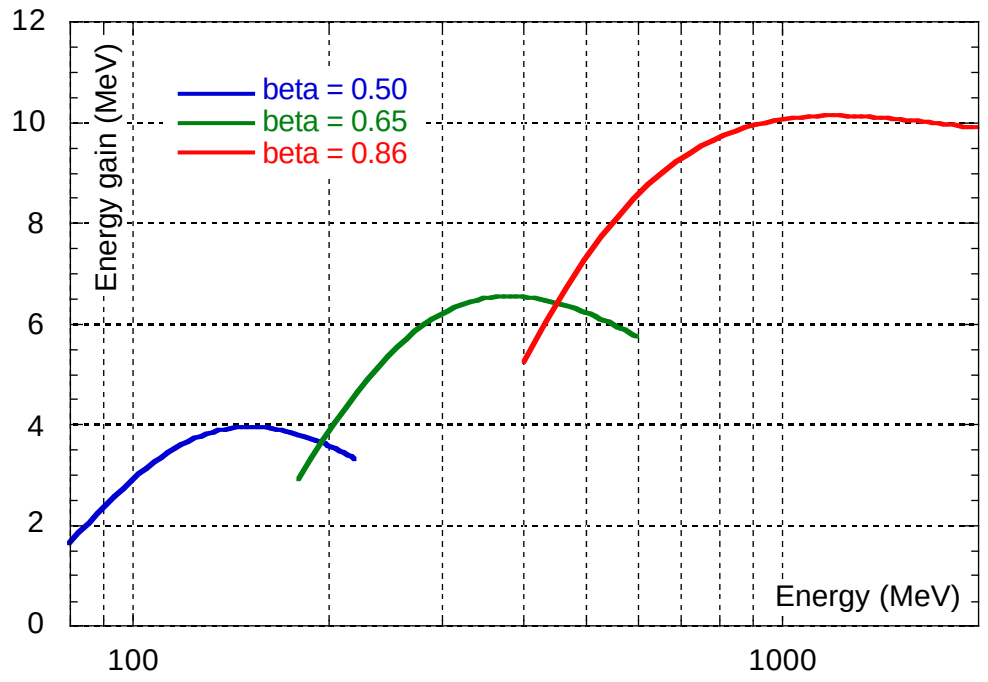

Figure 1: Energy gain for the 3 cavity types.

The operating accelerating fields G (net energy gain of a particle of constant speed equal to the geometric beta of the cell divided by the iris-to-iris cavity length) correspond to electric and magnetic peak surface fields of about 27 MV/m and 50 mT. The main characteristics of the 3 sectors are shown in Table 2. Longitudinal beam matching between sectors is controlled by adjustment of the synchronous phases of the two interface cryomodules. The resulting zero current phase advance per unit length ranges from about 20° to 4°. The beam power per cavity ranges from 130 kW at the beginning to 680 kW at the end of the linac for the mean beam current of 64.2 mA.

Table 2: Characteristics of the 3 sectors

| | Low-β | Medium-β | High-β |
|---|---|---|---|
| G (MV/m) | 8.5 | 10.5 | 12.5 |
| Geometric β | 0.5 | 0.65 | 0.86 |
| # cells | 5 | 5 | 5 |
| # cavities /cryom | 2 | 3 | 4 |
| # cryomodules | 16 | 14 | 23 |
| Sync phase (deg) | - 30 | - 27 | - 25 |
| Energy (MeV) | 85 - 195 | 195 - 450 | 450 - 1348 |

## 2 TRANSIENT BEAM-LOADING

Fig.2 shows for example the energy gain and the phase slippage with respect to a constant velocity particle,

which would stay on-crest of the RF wave, along the first cavity of the linac. With an accelerating gradient of 8.5 MV/m, the net energy gain is 2 MeV and the integrated beam phase is -30°.

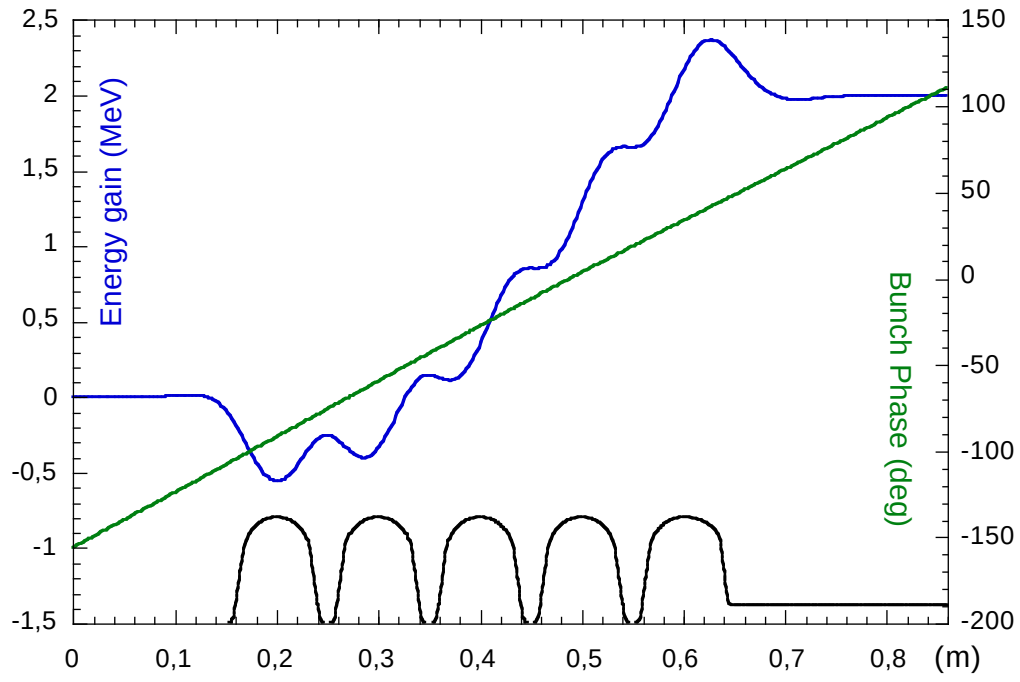

Figure 2: Energy gain and phase slippage along the $\beta$=0.50 cavity (G = 8.5 MV/m, $E_{in}$ = 85 MeV, $\phi_b$ = -30°).

This large phase slippage leads to different energy gains and then beam induced voltages between the cells of the cavity. As a result, very large fluctuations of transient beam-loading would be expected from multi-bunch trains. Fortunately, thanks to intercell coupling (about 1% here) RF power propagates from cell to cell and will tend to even the individual cell excitations. Some fluctuation however still remains due to the finite propagation velocity of the RF wave. In addition, with a chopped beam pulse, periodic gaps are cut in the regular bunch train. Modelling the cavity as a single resonator results in a perfect sawtooth-like voltage due to the periodic beam-loading and refilling of the cavity. With the multi-mode analysis, the fluctuation is increased and follows the oscillation caused by the closest mode to the accelerating Pi-mode of the passband. The energy gain modulation is then increased by a factor 3. (Fig.3).

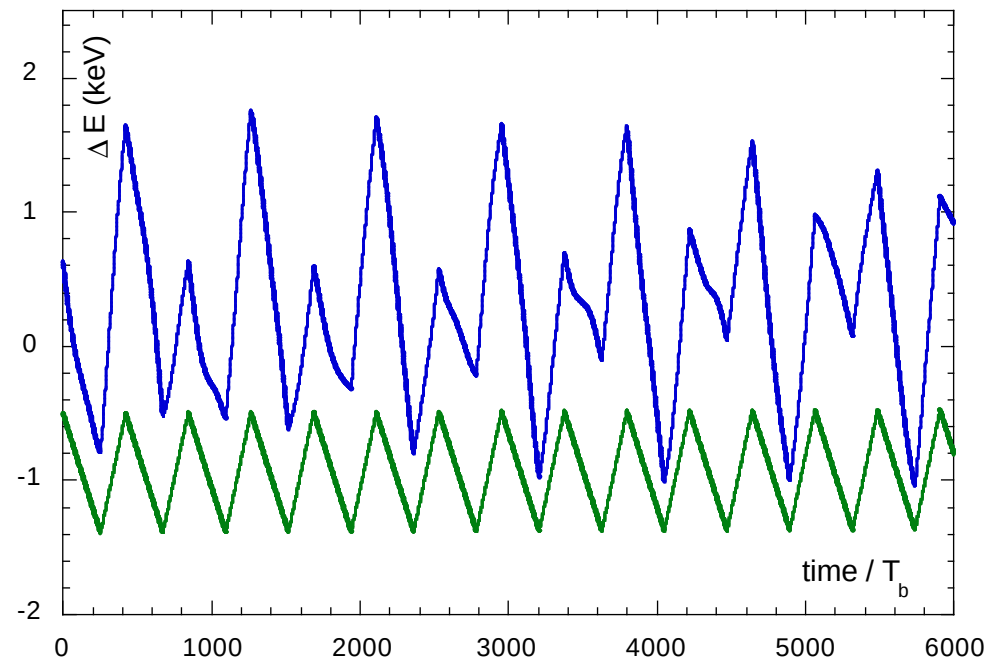

Figure 3: Bunch energy gain along - chopped beam pulse (bottom: single resonator model; top: multi-mode model).

## 3 CAVITY FIELD FLUCTUATIONS

Due to beam phase slippage effects of proton beams inside and outside the cavities, energy spread exhibits a much larger sensitivity to cavity fields fluctuations than for relativistic electron beams. Furthermore, in case of multiple cavities driven by one single power source, the control of the vector sum of the cavity voltages is not any more as efficient as for relativistic beams since the dynamic behaviour differs from one cavity to another. With N cavities driven by one single klystron, a total of 6×N coupled differential equations per klystron is required.

*- 3 equations per cavity for beam-cavity interaction*

Instead of using the crude RF-gap approximation (cosine-like acceleration at the cavity middle, corrected by the transit time factor) we preferred to integrate the exact differential equations in each cavity in order to model properly the beam-cavity interaction. Once the linac configuration has been defined (cavity types, number of cryomodules, design accelerating field and synchronous phase) a reference particle is launched through the linac in order to set the nominal phase of the field with respect to bunch at the entrance of all cavities.

*- 3×N equations per klystron for cavity field*

The dynamics of each resonator is described by two first order differential equations, plus another one modelling dynamic cavity detuning by the Lorentz forces [4]. Beam-loading is modelled by a cavity voltage drop during each bunch passage with a magnitude varying from cavity to another. In order to minimize the needed RF power,

1) the Qex is set near the optimal coupling (about 5 $10^5$)
2) the cavity is detuned to compensate the reactive beam-loading due to the non-zero beam phase

### *3.4 One cavity per klystron*

The cavity voltages are controlled by modulating the amplitude and phase of the power source via an I/Q modulator. The in-phase and out-of-phase feedback loop gains are set to 100 and 50, respectively. Examples of error sources, which were studied, are reproduced below.

*→ Input energy offset*

Bunch phase oscillations induced by beam injection errors will upset cavity voltage via beam-loading. Without feedback, the bunches become unstable very soon, while with feedback, cavity voltages are efficiently controlled and the constant field dynamics are recovered (Fig. 4).

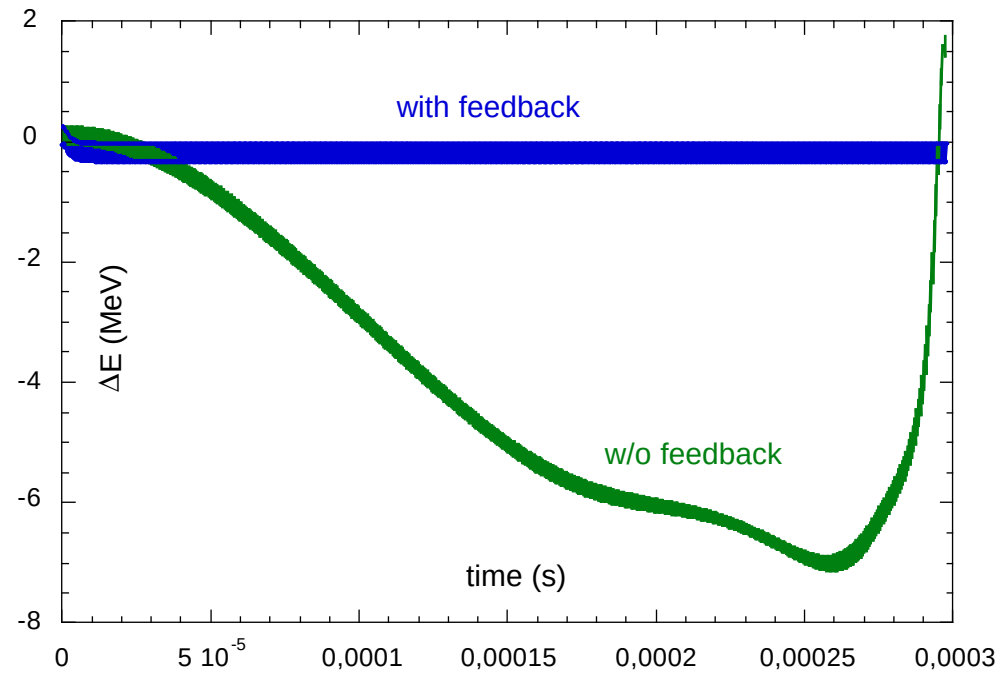

Figure 4: Beam energy deviation at linac exit without and with feedback (input energy error 0.5 %).

*→ Lorentz forces*

Because of the peak surface fields and of the mechanical rigidity of the structure, Lorentz force parameter increases as the cavity beta decreases. Simulations were carried out with expected values of 16, 8 and 4 $Hz/(MV/m)^2$ for the three $\beta$= 0.5, 0.65 and 0.86 cavity types. In order to relax the feedback requirements, the cavity must be pre-detuned, such that the resonance frequency equals the operating frequency at approximately half the beam pulse. The total detuning must then be set to the sum of the detunings for Lorentz forces and beam-loading compensations. Fig.5 shows the resulting phase and energy errors of all bunches of the train at the linac exit. The extra power is maximum at the low energy part of the linac (14%) and decreases as energy gain per cavity grows.

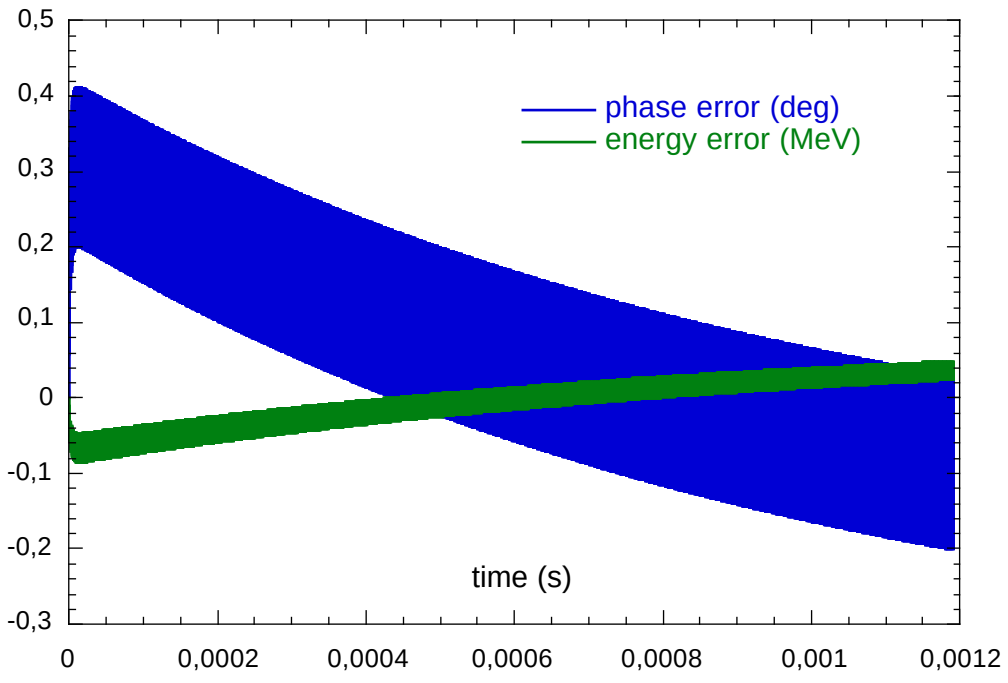

Figure 5: Bunch phase and energy deviations at linac exit (with Lorentz forces only).

→ *Microphonics*

With mechanical vibrations, feedback loops must be closed during the filling time, following pre-determined amplitude and phase laws, to ensure minimum RF power during the beam pulse. Assuming typical 40 Hz mechanical oscillation with an amplitude of 100 Hz (equivalent to phase fluctuations of ±8°), the increase in energy deviation at linac end is about 50% (Fig.6) and the extra peak power to be paid is about 20% at low energy.

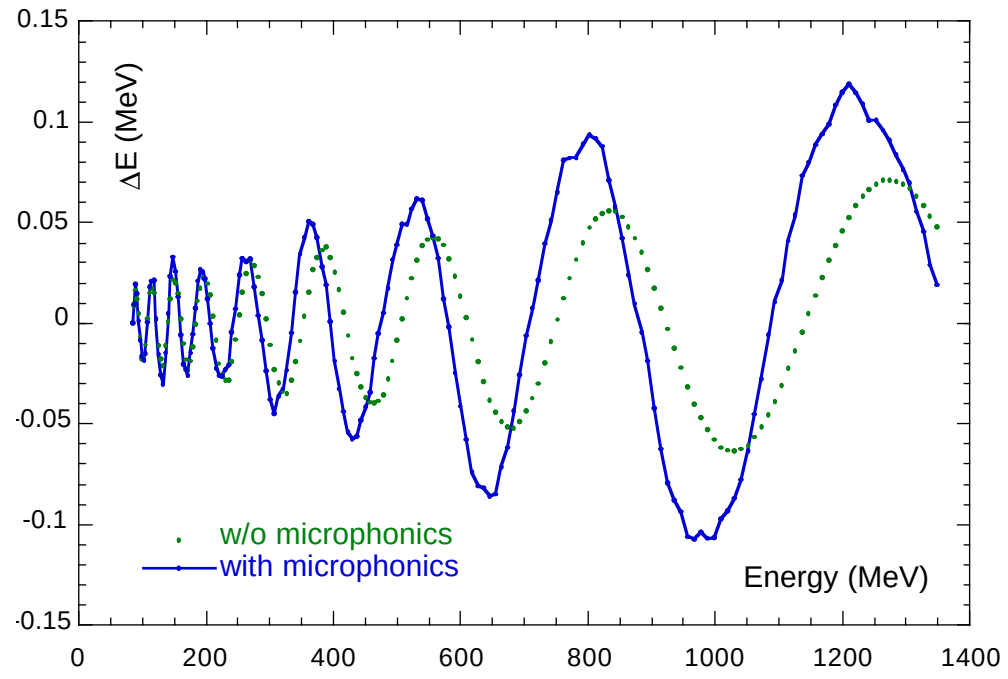

Figure 6: Energy deviation of last bunch along the linac (with Lorentz forces and microphonics).

## *3.5 Multiple cavities per klystron*

With relativistic electron beams, multiple cavities powered by a single power source can be easily controlled by the vector sum of the cavity voltages [4,5]. However for proton beams, since the dynamic behaviour of low-β cavities varies as the energy increases, even when the vector sum is kept perfectly constant, the individual cavity voltages can fluctuate dramatically. We could however envisage to feed individually the cavities at the low energy part of the SC linac and to feed groups of cavities by single klystrons at the high energy part, where the cavities have closer dynamic properties. Assuming for example groups of 8 cavities from the beginning of the 2nd sector only (above 200 MeV), Fig. 7 shows the 8 cavity voltages of the last klystron during the beam pulse with Lorentz forces detuning effects. The total energy deviation at linac end is lower than 1%.

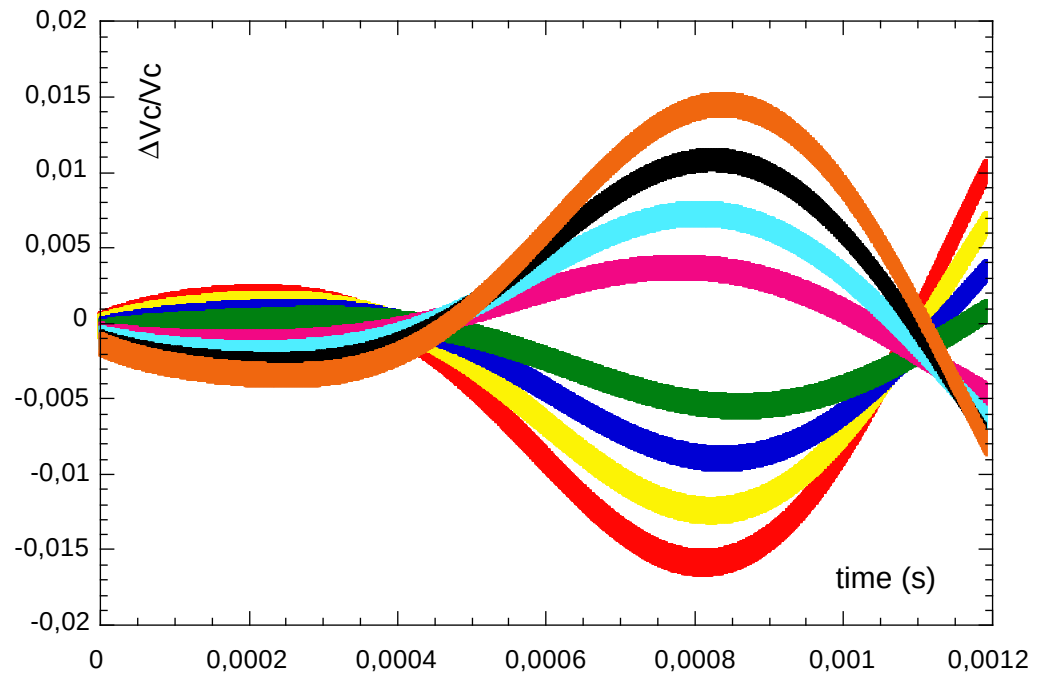

Figure 7: Amplitude of the 8 cavity voltages for the last group (with Lorentz forces only).

## 4 CONCLUSION

The systematic energy modulation, enhanced by transient beam-loading effects within a multi-cell cavity, looks actually harmless (about $10^{-3}$) by using SC cavities with low number of cells and not too small intercell coupling (respectively 5 and 1% in this study). Besides, the impact of various error sources on energy stability in a typical SC proton linac has been studied by means of a simulation code, which integrates the coupled differential equations governing both cavity field and beam-cavity interaction. When each cavity has its own RF feedback system, the cavity voltages can be very well controlled, providing energy spreads at the linac end well below the specifications. However, groups of multiple cavities driven by one common klystron and controlled by the vector sum, give rise to significant energy fluctuations and should only be used in high energy part, when the cavity dynamic properties become similar.